\documentstyle[12pt,aasms4,psfig]{article}
\slugcomment{original version submitted to Astronomical Journal: \quad {Jan 24, 2003 } }
\slugcomment{accepted for publication in Astronomical Journal: \quad {Jul 9, 2003 } }

\begin{document}

\hyphenation {non-re-la-ti-vi-stic}
\def\iacen{\'{\i}}

\title{Companions of Bright Barred Shapley Ames Galaxies}

\author{J. A. Garc\'{\i}a-Barreto\altaffilmark{1}, R. Carrillo\altaffilmark{1} }
\altaffiltext{1}{Instituto de Astronom\'{\i}a,
Universidad Nacional Aut\'onoma de M\'exico, Apdo Postal 70-264, M\'exico D.F.
04510, M\'exico; tony@astroscu.unam.mx, rene@astroscu.unam.mx}

\begin{center}
and
\end{center}

\author{N. Vera-Villamizar\altaffilmark{2} }
\altaffiltext{2}{Instituto Nacional de Astrof\'{\i}sica, Optica y Electr\'onica,
Calle Luis E. Erro 1, Tonanzintla, Edo. de Puebla, M\'exico, C.P. 72840
nelson@inaoep.mx}

\begin{abstract}
\noindent
	Companion galaxy environment for a subset of 78 bright and nearby barred galaxies from the
Shapley Ames Catalog is presented. Among the spiral barred galaxies there are Seyfert galaxies,
galaxies with circumnuclear structures, galaxies not associated with any large scale galaxy cloud
structure, galaxies with peculiar disk morphology (crooked arms) and galaxies with normal disk
morphology; the list includes all Hubbles types. The companion galaxy list includes the number of
companion galaxies within 20 diameters, their Hubble type and projected separation distance.
Additionally, companion environment was searched for four known active spiral galaxies, three of
them are Seyfert galaxies, namely, NGC 1068, NGC 1097, NGC 5548 and one is a starburst galaxy, M82.
Among the results obtained it is noted that the only spiral barred galaxy classified as Sy 1 in our
list has no companions within a projected distance of 20 diameters; 6 out of 10 Sy 2 bar galaxies
have no companions within 10 diameters, 6 out of 10 Sy 2 galaxies have one or more companions at
projected separation distances between 10 and 20 diameters; 6 out of 12 galaxies with
circumnuclear structures have 2 or more companions within 20 diameters.

\end{abstract}

\section{INTRODUCTION}

	Nearby barred spiral galaxies are important to study because they have a large scale stellar
bar and are better resolved compared with barred galaxies which are small and distant.
A non-axisymmetric gravitational potential can drive gas to flow from large to small radii
(\cite{com88}), yet not all barred galaxies present a similar level of central activity.
Bipolar nuclear outflows (as detected by their radio continuum emission) and circumnuclear structures
are perhaps a result of gas inflow; in particular, circumnuclear structures are thought to be a result
of a dynamical resonance between the pattern speed of the stellar bar and the orbital motion of gas
and stars in the disk, generally known as Inner Lindblad Resonance (\cite{bin87,but96}).

	The connection between nuclear activity and the properties of the parent galaxy is still
under investigation. The level of central (or nuclear) activity in galaxies is thought, in general,
to be mainly due to: a) merger events, (ie. ellipticals - radio galaxies [\cite{too72,too77}], ultra
luminous infrared galaxies [\cite{san96}]); b) tidal effects, (e.g., similar size nearby companions,
[\cite{sal91,nog88}]); c) non-axisymmetric gravitational potential (e.g., stellar bars);
d) minor mergers (with small galaxy satellites, [\cite{wal96}]). The central activity in spiral
galaxies is thought to be mainly due to the last three mechanisms.

	The issue of the role of bars may play on this process has been the subject of several
investigations (\cite{com88,shl89,ath92,phi94,ath94,wad95,ho96,mul97,sak99,sak00,com00,com03}). However, a conclusive
link between bars and nuclear activity has not yet been established (\cite{com00,com03}). One way to
further investigate this possible link is to focus not on how frequently active spiral galaxies
are barred, but on studying the properties of barred galaxies: for example, the strength of
the stellar bar, the frequency of companions, the atomic and molecular gas content, the group
membership and the possible existence of intergalactic gas capable of gas stripping as the barred
galaxy travels, the radio continuum bipolar outflows, etc. In this paper we concentrate on the
frequency of companions to barred galaxies. This would tell us if tidal interactions are
important with respect to the issue of activity in barred galaxies or if there are other factors
involved.

	Seyfert galaxies are found to be barred spirals of 79\% in the CfA sample
(\cite{kna00}) which suggests that a non-axisymmetrical gravitational potential might be relevant
to the activity. From statistical analysis the issue of large-scale companions near Seyfert
galaxies is still controversial but different studies indicate that Seyfert 1 galaxies do not
have an excess of companion galaxies, while Seyfert 2 galaxies have a
significant excess of large companions within a radius of 100 kpc (\cite{lau95,dul99}). This
result suggests that tidal interaction with a galaxy of similar size may induce central activity;
tails, as in the best example of tidal interaction between galaxies of similar size, are observed
in the Antennae and M51 systems (\cite{too72}). Observationally, it is very difficult to detect
companions of small dimensions and mass compared to the parent galaxy (\cite{iba94,iba97}); there
are, however, observations of several galaxies suggesting that the observed morphology is the
result of a merging event of a large (parent) galaxy and a minor galaxy in dimensions and mass
(e.g.,  M 101 [\cite{wal97}]; NGC 7479 [\cite{lai99}]; NGC 2110 [\cite{gon02}]).

	In this paper, we ask the following question: How many galaxy companions are there within
a distance of 10 and between 10 and 20 diameters (D$_{25}$)$\footnote{A projected distance of 20 D$_{25}$
diameters from a given barred galaxy is somewhat arbitrary; it is a large enough distance in order
to have an estimate of companions: for a galaxy with D$_{25}=3'$ at 40 Mpc, 20 D$_{25}$ would be
$\sim700$ kpc. Bergvall, Laurikainen \& Aalto [2003] consider an isolated galaxy as a galaxy
having no neighbors (with magnitude difference $\leq2$ magnitudes) closer than 6 diameters and
having no more than 2 neighbors within 16 diameters. Tully[1987] considers an isolated galaxy if
there are no companions within 6 Mpc.}$ of a given barred galaxy? What Hubble type is each companion galaxy?
At what projected distance is the companion galaxy from the barred galaxy under study? For this purpose we have searched for
companions, extracted from NED$\footnote{This research has made use of the NASA/IPAC
Extragalactic Database (NED) which is operated by the Jet Propulsion Laboratory, California Institute
of Technology, under contract with the National Aeronautic and Space Administration}$ with
the following criteria: a) the first and strongest criterium is that their distance was chosen
equal or less than 20 diameters from each galaxy in our study; b) for large-scale clouds of
galaxies, the systemic velocity of a companion should not differ from that of the barred galaxy
in question by more than twice the measured velocity dispersion of the cloud. If the cloud has no
measured velocity dispersion, then the velocity difference needs to be less than 300 km s$^{-1}$ as
discussed by Tully (1987); c) third, no attention was given to the visible magnitude of the companion.
The sample of 78 barred spirals, from the Shapley Ames Catalog, includes 2 SB0, 9 SBa, 1 SBab, 17 SBb,
17 SBbc, 22 SBc, 2 SBd, 3 SBm, 1 SB(r), and 1 SB(late). Of them 1 is Seyfert 1, 10 are Seyfert 2, 1
is Seyfert 2-like, 7 have circumnuclear structures, 11 have peculiar disk morphology (e.g., crooked
arms [\cite{gar91}]), 10 galaxies are not associated with any large scale galaxy sub structure
(according to Tully[1987,1988=T88]) and 38 have normal disk morphology; they all are in the distance
interval 2.2 Mpc $\leq D \leq 53.2$ Mpc. This subset of barred galaxies is not complete to any limiting distance or apparent magnitude, but it is a
representative subset of nearby barred spiral galaxies with different Hubble types and members of
different cloud or group associations. This sample was originally chosen based on far
infrared (IRAS) colors (log[f12/f25]$\leq-0.15$ and log[f60/f100]$\leq-0.1$) indicative of central
star formation (after \cite{hel86}) or relatively warm dust (IRAS) temperature, T$_d\geq29$ K,
(\cite{gar93,gar96}) at $\delta\geq-41$. The study of $''$bar galaxies and their environment$''$
by van den Bergh (2002) focused in assigning each of all northern Shapley Ames galaxies to either
field, group or cluster environment. We believe that our approach has the advantage over van den
Bergh's study by giving a list of names of companions, their Hubble type, systemic velocity and
projected distance from 78 Shapley Ames barred galaxies.

	Table 1 lists group membership and (cosmological) distances (using H$_0=75$ km s$^{-1}$
Mpc$^{-1}$, T88) and Tables 2 and 3 list the projected linear separations (based on angular distances
estimated from NED) to each nearest large scale neighbor galaxy (using systemic velocity information
from T88 and NED).

\subsection{Tully's Group Terminology}

	Since we refer to the group assignment of each galaxy according to T88, a brief description
of his terminology$\footnote{For a complete description of group terminology please refer to
Tully[1987]}$ is in order: a) a galaxy is part of a group under the density
criterion based on luminosity and distance separation as $\rho=LR_{ij}^{-3}$, specifically
$\rho_g=2.5\times10^9 L_{\odot}Mpc^{-3}$; b) galaxies can be associated on two levels, i) at first
level there are entities that failed the criterion for a group but passed the criterion for
association,$\rho_g=2.5\times10^8 L_{\odot}Mpc^{-3}$; ii) at a second level, several groups or
first level associations of individual galaxies may link together at densities that satisfy the
definition of an association; c) a galaxy is assigned to a structure identified by
xx$\pm$yy$\pm$zz, where xx refers to a Cloud or Spur, -yy refers to a group, +yy refers to a first
level association and +zz refers to a second level association; d) a galaxy identified with a
cloud that is not associated with any substructure would have group assignment xx-0 (ie. NGC 2525
has group assignment 31-0 in the Antlia-Hydra Cloud); groups and associations within a cloud are
numbered in order according to their proximity to the prominent structure within a cloud
(e.g., NGC 4477 has a group assignment 11-1 which means it is within the Virgo Cluster, while
NGC 1415 has a group assignment 51-4+4 which means that it is a little away from the Fornax
Cluster [$\sim4$ Mpc away]; e) if a galaxy in the general region X cannot be associated with any cloud, then it has the group
assignment x0$\pm$yy where yy=-0 is given if the galaxy is not in a group or association
(e.g., NGC 6951 has group assignment 40-0 in the Leo Clouds); f) a galaxy is called isolated if it
is farther than 6 Mpc from any high probability member.

	Table 1 lists the barred galaxies in our study; the table is divided in five groups, namely,
galaxies with Seyfert activity, galaxies having circumnuclear structures, galaxies with peculiar disk
morphology, galaxies not associated with any large scale (cloud) substructure as defined by T88 and
galaxies with normal disk morphology. Table 2 lists the number of companions extracted from NED
according to angular separation and systemic velocity (see below). Table 3 lists a summary of Table 2
with number of companions from every barred galaxy in our study within 10 and between 10
and 20 parent galaxy diameters.

\section{RESULTS}

	 Gas flow is a necessary condition for nuclear or central activity, a condition which
could be satisfied by a large-scale stellar bar. However evidence linking actual bars to nuclear
activity has been weak(\cite{sak99,com00,sak00,com03,mar03}).
Observational evidence of (stellar and gas) long tails or optical morphology deformation as of the type of
the antennae was shown to be the result of tidal interaction or a recent major merger event
(\cite{too72,rot90,tys98,sal99a,sal99b,hib00}). Numerical simulations indicate that merging of two
spiral disks of equal mass might result in a completely different stellar population, for example a
galaxy with shells (\cite{her92}). Long exposure optical images of the spiral galaxy NGC 5548 (SA0/a)
classified as Seyfert 1.5 shows a long, low surface brightness structure suggesting that its origin
could have been the result of a major merger (\cite{tys98}). Only one of the galaxies in our study
presents a large scale long HI tail (\cite{hog01}), namely NGC 6239, a galaxy not associated with any
large scale (cloud) substructure, and no large scale companion within 20 diameters.

	Our study identifies companion galaxies only within certain projected distances and
having similar systemic velocities as each barred galaxy in our list; true spatial
separations remain to be estimated as distances to each individual barred galaxy and companion are
determined (see below for M82 system).

	Tables 2 and 3 indicate that the one Sy 1 galaxy, NGC 3783, has no
large scale companions within a distance of 20 diameters (38$'$ [425 kpc]); 6 out of 10 Sy 2
galaxies have no companions within 10 diameters; 7/10 Sy 2 galaxies have one or no companions
and 3/10 Sy 2 galaxies have three or more companions within a separation distance between 10
and 20 diameters; 3 Sy 2 galaxies show circumnuclear structures with no companions within 10
diameters, 1 Sy 2 galaxy has one companions at a distance between 10 and 20 diameters 1 Sy 2
galaxy has four companions at a distance between 10 and 20 diameters. The only Sy 2-like galaxy has
no companions within 10 diameters and one companion within a separation distance between 10 and 20
diameters and yet it presents a bipolar radio continuum emission with large scale (12 kpc in diameter)
lobes straddling the nucleus (\cite{gar98,gar02}). From the group of galaxies with circumnuclear
structures (CNS), 6 out of 12 have 1 or more companions at distances within 10 diameters and 6/12
have 2 or more companions within a distance between 10 and 20 diameters. From the group of galaxies
with peculiar disk morphology (PM), 7/14 have no companions, 4/14 have one companion within 10
diameters; 11/14 have one or no companions and 3/14 have two or more companions within a distance
between 10 and 20 diameters. From the group of galaxies not associated with any large scale (cloud)
[NA] substructure 10/13 have no companions within 10 diameters, 9/13 have no companions and 4/13 have
1 or more companions within a distance between 10 and 20 diameters. From the group of galaxies with
normal disk morphology (ND), 11/38 have no companions, 13/38 have 1 companion and 14/38 have 2 or
more companions within 10 diameters, 12/38 have no companions, 10/38 have 1 companion and 16/38 have
2 or more companions within a separation distance between 10 and 20 diameters. Figure 1 shows two
histograms one with number of Seyfert 2 galaxies versus number of companions, the other, with galaxies
with circumnuclear structures versus number of companions. Figure 2 shows the corresponding histograms
for galaxies with peculiar morphology and normal disk morphology. Finally Figure 3 shows the
histogram of galaxies not associated with any large scale cloud structure versus the number of
companions.

	As seen in Table 3 the galaxies with more companions within 20 diameters are NGC 4435 (SB0),
NGC 4535 (SBc) and NGC 4654 (SBc) which are in the Virgo cluster of galaxies and present {\it normal}
disk morphology. The galaxies having circumnuclear structures NGC 1326, NGC 1415, NGC 3351, and
NGC 4314, have 7 or more companions within 20 diameters; each of them belongs to a group of galaxies.

	As an extension (check) of the method used for finding the companions of barred galaxies we have
done a similar search for 4 known spiral galaxies that exhibit different central activity; these
galaxies are NGC 1068, NGC 1097, M82 and NGC 5548. These galaxies are listed in Table 4. Companions
to these galaxies are listed in Table 5 and a summary of companions to these galaxies within a
separation distance less than 10 diameters and between 10 and 20 diameters is in Table 6. NGC 5548
type (R')SA(s)0/a, X-ray emmitter and Sy 1.5 galaxy has no companions within 10 diameters and 3 at projected
separations larger than 300 kpc; NGC 1068, Sy 1, has one companion at 43 kpc (10$'.2$) and two
others at projected distances more than 300 kpc; NGC 1097, Sy 1, has one companion at 14 kpc
(3$'.4$) and two others at 195 and 303 kpc; M82, a starburst galaxy, has 6 companions within 10
diameters and 7 more at projected separations from 121 kpc to 204 kpc.

	All companions found for our list of barred galaxies are considering projected distances
but one must be aware that true spatial distance to companions remain to be determined on a case
by case basis; M82 system is a good example when one is considering the question: is the projected separation
distance representative of the true spatial distance separation? M82 has, among others, the
companions M81 and NGC 3077 at projected separations of only 39 kpc and 73 kpc respectively assuming
that M82 is at the cosmological distance of 3.6 Mpc as M81; however, M82 and NGC 3077 previous
cosmological distance were estimated to be 5.2 Mpc and 2.1 Mpc, respectively(T88). The detection of
HI in the system M81, M82, NGC 3077 (\cite{yun94}) supports the idea that the three galaxies are at
the same (cosmological) distance as M81 as determined by HST Cepheids (\cite{fre94}); true spatial
distances, though, remain to be estimated after cosmological distances to M82 and NGC 3077 are
determined individually.

\section{CONCLUSIONS}

	We have made a detailed study of likely physical companions to a sample of 78 nearby
barred spirals, initially chosen on the basis of IRAS properties and fairly representative of
bright barred spirals. We do not find a strong correlation between the presence of a
companion within 10 or 20 projected diameters and nuclear activity in these barred galaxies.
Histograms indicate a wide range of companion numbers for each of our subcategories of barred
galaxies. Nuclear activity (as Seyfert type) in barred galaxies is not necessarily due to an
enhanced presence of companions to these galaxies, a result which is in agreement with other
studies indicating that for Seyfert 1 galaxies there is no clear evidence of any excess of
companion galaxies within 100 kpc or within 3 diameters (\cite{dul99}); in fact, the only Seyfert
1 galaxy in our study, namely, NGC 3783, has no companions within a projected separation of 20
diameters or 425 kpc. Six out of ten Seyfert 2 galaxies (Seyfert 2 galaxies are thought to have
companions within a search radius of 100 kpc [\cite{dul99}]), in our study, have no companions
within a projected distance of 10 diameters, three barred Sy 2 galaxies have no companions between 10
and 20 diameters, two barred Sy 2 galaxies have one companion between a distance of 10 and 20
diameters, and one barred Sy 2 galaxy has two companions between 10 and 20 diameters. Finally our
study suggests that having a large-scale stellar bar and nearby companions are not sufficient for
a spiral galaxy to present nuclear activity: at present it is beyond the scope of this paper to
explain why barred galaxies not associated with any large scale cloud of galaxies and therefore no
companions do not present nuclear activity as does the only Seyfert 1 in our study, NGC 3783 (which
has no companions); or for that matter explain why NGC 4435 or NGC 4535, two barred galaxies with
the most number of companions do not present any nuclear activity (in case tidal interactions was
an important mechanism). In a forthcoming paper we will present major and minor diameters of
large scale stellar bars from 46 nearby bright barred galaxies; this could be another factor for
the difference in activity in barred spirals.

\section*{Acknowledgements}

	It is a pleasure to thank an anonymous referee for his(her) valuable comments to improve
this paper. We also thank E. Moreno for his comments about this work.

\clearpage
\newpage
\begin{table}
\small
\caption[ ]{
Shapley Ames Barred Spiral Galaxies}
\begin{flushleft}
\begin{tabular}{llllr}
\hline
{\bf Galaxy}  & Hubble & Characteristic & Tully's   & Distance  \cr
{\bf NGC/IC} & Type    & $^a$& Group$^b$ & Mpc       \cr
\hline
N 3185 & SBa(s) & Sy 2 & 21-6+6 & 21.3 \cr
N 3367 & SBc(s) & Sy 2-like, CNS, PM & 32-4+4 & 43.6 \cr
N 3783 & SBa(r) & Sy 1 & 31+11+10 & 38.5 \cr
N 4477 & SBa & Sy 2 & 11-1 & 16.8 \cr
N 4507 & SBab(rs) & Sy 2 & - & 45.0 \cr
N 4725 & SBb(r) & Sy 2 & 14-2+1 & 12.4 \cr
N 5135 & SBb & Sy 2, CNS & - & 53.2 \cr
N 5347 & SBb(s) & Sy 2, CNS, NA & 42-0+1 & 36.7  \cr
N 5728 & SBb(s) & Sy 2, CNS & 41+15+15 & 42.2  \cr
N 6217  & RSBbc(s) & Sy 2, NA & 44-0+5 & 23.9 \cr
N 6951 & SBb(rs) & Sy 2, CNS, NA & 40-0 & 24.1 \cr
N 7479 & SBbc(s) & Sy 2 & 64-2+1 & 32.4  \cr
\hline
N 1022 & SBa(r) & CNS & 52-1+1 & 18.5 \cr
N 1326 & SBa & CNS & 51-1+1 & 16.9 \cr
N 1415 & SBa & CNS & 51-4+4 & 17.7 \cr
N 3318 & SBbc & CNS & 31-6+6 & 37.9  \cr
N 3351 & SBb & CNS & 15-1+1 & 8.1 \cr
N 4314 & SBa & CNS & 14-1+1 & 10.0  \cr
N 5430 & SBb & CNS & - & 38.0 \cr
\hline
N 3319  & SBc(s) & PM & 15+7 & 11.5 \cr
N 1637  & SBc & PM & 53-20+20 & 8.9$^c$ \cr
N 2139  & SBc(s) & PM & 34+1 & 22.4 \cr
N 2798  & SBa(s) & PM & 21-6+16 & 27.1 \cr
\end{tabular}
\end{flushleft}
\end{table}
\clearpage
\newpage
\setcounter{table}{0}
\begin{table}
\small
\caption[ ]{
Shapley Ames Barred Spiral Galaxies}
\begin{flushleft}
\begin{tabular}{llllr}
\hline
{\bf Galaxy}  & Hubble & Characteristic & Tully's   & Distance  \cr
{\bf NGC/IC} & Type    & $^a$& Group$^b$ & Mpc       \cr
\hline
N 4618  & SBbc(rs) & PM & 14-4 & 7.3 \cr
N 5534  & SBbc & PM & - & 35.0\cr
N 5597  & SBc(s) & PM & 41-14+14 & 38.6 \cr
N 5691  & SBb    & PM & 41-2+1 & 30.2 \cr
N 5757  & SBb(rs) & PM & 41+15+15 & 39.5  \cr
N 5915  & SBbc(s) & PM & 41-10+10 & 33.7 \cr
N 6907  & SBbc & PM & - & 42.0 \cr
\hline
I 1953  & SBbc & NA, PM & 51-0+4 & 22.1\cr
N 2525  & SBc & NA, PM & 31-0 & 21.1 \cr
N 2787  & SBa(s) & NA & 12-0 & 13.0 \cr
N 3359  & SBc(s)pec & NA & 12-0+1 & 19.2 \cr
N 4561  & SBc & NA & 11-0+1 & 12.3 \cr
N 4691  & SB0 & NA & 11-0+10 & 22.5 \cr
N 5669  & SBc(r) & NA & 41-0+1 & 24.9 \cr
N 5921  & SBbc(s) & NA & 41-0+1 & 25.2 \cr
N 6239  & SBc pec & NA & 44-0 & 18.9 \cr
I 5273  & SBc & NA, PM & 61-0+16 & 16.0 \cr
\hline
N 672   & SBbc(s) & ND & 17-5+5 & 7.5 \cr
N 1784  & SBbc(r) & ND & 34+5+4 & 28.7 \cr
N 1832  & SBb(r)  & ND & 34-3+3 & 23.5 \cr
N 2217  & SBa(s)  & ND & 34-1+1 & 19.5 \cr
N 2223  & SBbc(r) & ND & 34-8+8 & 33.7 \cr
N 2336  & SBbc(r) & ND & 42-17+16 & 32.9 \cr
N 2339  & SBc(s)  & ND & 30+1 & 30.9 \cr
\end{tabular}
\end{flushleft}
\end{table}
\clearpage
\newpage
\setcounter{table}{0}
\begin{table}
\small
\caption[ ]{
Shapley Ames Barred Spiral Galaxies}
\begin{flushleft}
\begin{tabular}{llllr}
\hline
{\bf Galaxy } & Hubble & Characteristic & Tully's  & Distance \cr
{\bf NGC/IC} & Type & $^a$& Group$^b$ & Mpc  \cr
\hline
N 2366  & SBm IV  & ND & 14-10 & 2.9 \cr
N 2545  & SBbc(r) & ND & - & (45.1) \cr
N 2835  & SBc(rs) & ND & 54-3+1 & 10.8 \cr
N 2935  & SBb(s)  & ND & 3-12 & 30.6 \cr
N 3287  & SBc(s)  & ND & 21-6 & 20.6 \cr
N 3504  & SBb(s)  & ND & 21-7+7 & 26.5 \cr
N 3513  & SBc(s)  & ND & 54-5+5 & 17.0 \cr
N 3686  & SBc(s)  & ND & 21-1 & 23.5 \cr
N 3729  & SB(r)   & ND & 12-1 & 17.0 \cr
N 3912  & SB(late) & ND & 13-9+9 & 30.0 \cr
N 3953  & SBbc(r) & ND & 12-1 & 17.0 \cr
N 3992  & SBb(rs) & ND & 12-1 & 17.0 \cr
N 4123  & SBbc(rs) & ND & 22-11+11 & 25.3 \cr
N 4214  & SBm III & ND & 14-7 & 3.5 \cr
N 4236  & SBd IV  & ND & 14+10 & 2.2 \cr
N 4242  & SBd III & ND & 14-4 & 7.5 \cr
N 4385  & SBbc(s) & ND & 11-25+24 & 33.5 \cr
N 4435  & SB0     & ND & 11-1 & 16.8 \cr
N 4487  & SBc(s)  & ND & 11-14+10 & 19.9 \cr
N 4496A & SBm(rs) & ND & 11-4+1 & 13.1 \cr
N 4535  & SBc(s)  & ND & 11-1 & 16.8 \cr
N 4654  & SBc(rs) & ND & 11-1 & 16.8 \cr
N 4688  & SBc(s)  & ND & 11+2+1 & 17.1 \cr
N 4902  & SBb(s)  & ND & 11-30 & 39.2 \cr
\end{tabular}
\end{flushleft}
\end{table}
\clearpage
\newpage
\setcounter{table}{0}
\begin{table}
\small
\caption[ ]{
Shapley Ames Barred Spiral Galaxies}
\begin{flushleft}
\begin{tabular}{llllr}
\hline
{\bf Galaxy } & Hubble & Characteristic & Tully's  & Distance \cr
{\bf NGC/IC} & Type & $^a$& Group$^b$ & Mpc  \cr
\hline
N 4981  & SBbc(rs) & ND & 11-17+10 & 27.8 \cr
N 5068  & SBc(s)  & ND & 14+17 & 6.7 \cr
N 5792  & SBb(sr) & ND & 41+2+1 & 30.6 \cr
N 5850  & SBb(sr) & ND & 41-1 & 28.5 \cr
N 7640  & SBc(s)  & ND & 65-4 & 8.4 \cr
N 7723  & SBb(rs) & ND & 63-6 & 23.7 \cr
N 7741  & SBc(s)  & ND & 65-3 & 12.3 \cr
\end{tabular}
\end{flushleft}
$^a$ Main Characteristic: Sy: Seyfert type; CNS: circumnuclear structure; PM: peculiar disk
morphology; NA: galaxies not associated with any substructure, according to Tully(1987,1988)[refer to
section 1 for a brief description when a galaxy is associated to a group or spur]; ND : normal disk.

$^b$ T88; first digit specifies one of seven regions in T88, the second digit identifies
a specific cloud within the general region; for a complete description refer to section 1 of this
paper and Tully(1987) 11: Virgo cluster; 12: Ursa Major cloud; 13: Ursa Major Southern Spur;
14: Coma I Sculptur cloud; 15: Leo Spur; 17: Triangulum Spur; 21: Leo cloud; 22: Crater cloud;
31: Antlia-Hydra cloud; 32: Cancer Leo cloud; 34: Lepus Cloud; 40: Leo clouds; 41: Virgo Libra cloud;
42: Canes Venatici Camelopardalis cloud; 44: Draco cloud; 51: Fornax cluster and Eridanus cloud;
52: Cetus-Aries cloud; 53: Dorado cloud; 61: Telescopium Grus cloud; 63: Piscis Austrinus cloud;
64: Pegasus cloud.

$^c$ New Cepheid distance has been determined to be 11.7 Mpc (\cite{leo03}). This new distance does
not modify any of the analysis in this paper.

\end{table}
\clearpage
\newpage

\begin{table}
\small
\caption[ ]{
Large Scale Galaxy Nearest Neighbors$^a$}
\begin{flushleft}
\begin{tabular}{llrrllrr}
\hline
{\bf Galaxy} & Characteristic & V$_{sys}^b$ & V$_{\sigma}^c$ & Companion$^d$  & Hubble & V$^e_{sys}$ & Separation $^f$ \cr
{\bf NGC/IC} & & km s$^{-1}$ & km s$^{-1}$ &  & type & km s$^{-1}$ & kpc \cr
\hline
N 3185 & Sy 2 & 1237 & 124 & N3190 & Sa & 1271 & 66 \cr
         & & & &N3187 & SBc(s) & 1521 & 70 \cr
	 & & & &N3193 & E2 & 1399 & 98 \cr
	 & & & &N3177 & SAb(rs) & 1302 & 230 \cr
N 3367 & Sy 2-like, CNS, PM & 3045 & 0 & N3391 & S & 2956 & 563 \cr
N 3783 & Sy 1 & 2910 & & &  & & \cr
N 4477 & Sy 2 & 1263 & 715  & N4479 & SB0(s) & 876 & 25 \cr
         & & & & N4473 & E5 & 2244 & 62 \cr
         & & & & N4474 & S0 & 1610 & 127 \cr
	 & & & & N4461 & SB0(s) & 1931 & 150 \cr
N 4507 & Sy 2 & 3525 & & & & & \cr
N 4725 & Sy 2 & 1210 & - & N4747 & SBcd & 1190 & 87 \cr
         & & & & Mrk 1338 & E & 1069 & 141 \cr
	 & & & & KVG 1249+263 & Irr & 1225 & 146 \cr
	 & & & & N4670 & SB0/a & 1069 & 431 \cr
	 & & & & N4565 & Sb(s) & 1282 & 694 \cr
	 & & & & UGCA 294 & S0 pec & 947 & 701 \cr
	 & & & & N4562 & SBdm(s) & 1353 & 728 \cr
N 5135 & Sy 2, CNS & 4157 & & I4248 & E & 4133 & 218 \cr
         & & & & CEN77 09 & Irr & 4413 & 224 \cr
	 & & & & N5152 & SBb(s) & 4147 & 472 \cr
	 & & & & N5124 & E6 & 3976 & 476 \cr
	 & & & & N5153 & E1 & 4281 & 483 \cr
	 & & & & E444-G021 & Sc & 4265 & 515 \cr
	 & & & & E444-G030 & Sa & 4225 & 670 \cr
N 5347 & Sy 2, CNS, NA & 2386 & &  &  & & \cr
N 5728 & Sy 2, CNS & 2780 & & N5744 & Sa & 2692 & 1190 \cr

\hline
\end{tabular}
\end{flushleft}
\end{table}
\clearpage
\newpage
\setcounter{table}{1}
\begin{table}
\small
\caption[ ]{
Large Scale Galaxy Nearest Neighbors$^a$}
\begin{flushleft}
\begin{tabular}{llrrllrr}
\hline
{\bf Galaxy} & Characteristic & V$_{sys}^b$ & V$_{\sigma}^c$ & Companion$^d$ & Hubble & V$^e_{sys}$ & Separation $^f$ \cr
{\bf NGC/IC} & & km s$^{-1}$ & km s$^{-1}$ &  & type & km s$^{-1}$ & kpc \cr
\hline
N 6217 & Sy 2 & 1359 & - & & & & \cr
N 6951 & Sy 2, CNS, NA & 1426 & & & & &\cr
N 7479 & Sy 2 & 2384 & 134 & U12300 & SBa & 2294 & 650 \cr
         & & & & U12281 & Sd & 2568 & 1071 \cr
N 1022 & CNS & 1574 & 99 & N 961 & SBm(rs) & 1300 & 216 \cr
         & & & & UGCA 038 & SBm(s) & 1327 & 242 \cr
N 1326 & CNS & 1364 & 434 & N1326A & SBm(s) & 1836 & 77 \cr
         & & & & N1326B & SBm(s) & 1006 & 86 \cr
	 & & & & N1316C & RSA0 & 1800 & 172 \cr
	 & & & & N1317  & RSAB0(rl) & 1941 & 201 \cr
	 & & & & N1316  & RSAB0(s) & 1760 & 231 \cr
	 & & & & N1310  & SBcd(rs) & 1805 & 253 \cr
	 & & & & N1341  & SABab(rs) & 1813 & 312 \cr
N 1415 & CNS & 1553 & 110 & N1401 & SB0(s) & 1518 & 125 \cr
	 & & & & N1426 & E4 & 1443 & 193 \cr
	 & & & & N1395 & E2 & 1717 & 227 \cr
	 & & & & N1414 & SBbc(s) & 1681 & 262 \cr
	 & & & & N1422 & SBab & 1637 & 275 \cr
	 & & & & N1438 & SB0(r) & 1555 & 336 \cr
	 & & & & N1439 & E1 & 1670 & 340 \cr
N 3318 & CNS & 2910 & 52 & N3318B & SBc(s) & 2756 & 113 \cr
	 & & & & E318-G003 & Sc & 2767 & 521 \cr
N 3351 & CNS & 780  & 112& N3368 & Sab(s) & 897 & 99 \cr
         & & & & CGCG065-086 & dS0 & 778 & 112 \cr
	 & & & & N3379 & E0 & 911 & 182 \cr
	 & & & & 1046+1234 & & 887 & 186 \cr
	 & & & & N3384 & SB0(s) & 704 & 199 \cr
\hline
\end{tabular}
\end{flushleft}
\end{table}
\clearpage
\newpage
\setcounter{table}{1}
\begin{table}
\small
\caption[ ]{
Large Scale Galaxy Nearest Neighbors$^a$}
\begin{flushleft}
\begin{tabular}{llrrllrr}
\hline
{\bf Galaxy} & Characteristic & V$_{sys}^b$ & V$_{\sigma}^c$ & Companion$^d$ & Hubble & V$^e_{sys}$ & Separation $^f$ \cr
{\bf NGC/IC} & & km s$^{-1}$ & km s$^{-1}$ & & type  & km s$^{-1}$ & kpc \cr
\hline
	 & & & & N3299 & SABdm(s) & 641 & 297 \cr
	 & & & & N3412 & SB0(s) & 841 & 340 \cr
	 & & & & N3377 & E6 & 665 & 348 \cr
N 4314 & CNS & 980 & 266 & N4308 & E & 624 & 40 \cr
         & & & & U7457 & Sc & 659 & 101 \cr
	 & & & & N4274 & Sa(s) & 930 & 117 \cr
	 & & & & N4286 & SA0(r) & 644 & 122 \cr
	 & & & & N4310 & RSAB0(r) & 913 & 124 \cr
	 & & & & N4283 & E0 & 1076 & 136 \cr
	 & & & & N4278 & E1 & 649 & 145 \cr
	 & & & & I3247  & Sd & 569 & 182 \cr
	 & & & & N4245 & SBa(s) & 815 & 199 \cr
N 5430 & CNS & 2960 & & N5402 & S & 3021 & 385 \cr
\hline
I 1953 & PM, NA & 1859 & & E548-G034 & SB & 1664 & 163 \cr
	  & & & & I1962 & SBdm(s) & 1806 & 186 \cr
	  & & & & N1377 & S0 & 1792 & 346 \cr
N 1637  & PM    & 717  & 31 &  & &  &  \cr
N 2139  & PM    & 1843 &    &  & &  &  \cr
N 2525  & PM, NA & 1585 & & &  & & \cr
N 2798  & PM    & 1744 & 124 & N2799 & SBm(s) & 1865 & 12 \cr
        &       &      &     & KTG 22 & - & 1799 & 23\cr
	&       &      &     & UGC 04904 & SB & 1670 & 43 \cr
N 3319  & PM    & 744  &     &       &        &      & \cr
N 4618  & PM    & 544  & 58  & VV 073A & - & 562 & 2 \cr
        &       &      &     & N4625 & SABm(rs) & 609 & 17 \cr
	&       &      &     & UGC 07751 & Im & 605 & 153 \cr
\hline
\end{tabular}
\end{flushleft}
\end{table}
\clearpage
\newpage
\setcounter{table}{1}
\begin{table}
\small
\caption[ ]{
Large Scale Galaxy Nearest Neighbors$^a$}
\begin{flushleft}
\begin{tabular}{llrrllrr}
\hline
{\bf Galaxy} & Characteristic & V$_{sys}^b$ & V$_{\sigma}^c$ & Companion$^d$ & Hubble & V$^e_{sys}$ & Separation $^f$ \cr
{\bf NGC/IC} & & km s$^{-1}$ & km s$^{-1}$ & & type  & km s$^{-1}$ & kpc \cr
\hline
N 5534  & PM & 2633 & &  &  & & \cr
N 5597  & PM & 2683 & & N5595 & SABc(rs) & 2711 & 46 \cr
N 5691  & PM & 1870 & 136 & SD:J143$^g$ & Sd & 300 \cr
        &    &      &     & N5705 & SBd(rs) & 1757 & 303 \cr
	&    &      &     & N5713 & SAB(rs) & 1972 & 306 \cr
N 5757  & PM & 2678 & & E580-G034 & SBc & 2727 & 44\cr
        & & & & E580-G029 & Sb(r) & 2596 & 476 \cr
N 5915  & PM & 2274 & 90 & N5916 & SBa(rs) & 2302 & 43 \cr
	&    &      &    & N5916A & SBc(s) & 2292 & 45 \cr
N 6907  & PM & 3180 & & N6908 & S &  & 10$^h$ \cr
          & & & & I5005 & SBcd(s) & 3112 & 749 \cr
\hline
N 2787  & NA & 700  & & U4998 & Im & 623 & 222 \cr
N 3359  & NA & 1018 & & &  &  & \cr
N 4561  & NA & 1407 & & &  &  &  \cr
N 4691  & NA & 1098 & &  & & & \cr
N 5669  & NA & 1371 &  & KUG142 & BCD & 1382 & 44 \cr
N 5921  & NA & 1480 &  & N5921:[KSF97]B & & 1532 & 15 \cr
N 6239  & NA & 922 & &  &  & &  \cr
I 5273  & NA, PM & 1320 & & E406-G040 & Irr & 1248 & 150 \cr
	  & & & & E406-G042 & SABm(s) & 1375 & 232 \cr
	  & & & & N7418 & Sc(rs) & 1446 & 247 \cr
\hline
\end{tabular}
\end{flushleft}
\end{table}
\clearpage
\newpage
\setcounter{table}{1}
\begin{table}
\small
\caption[ ]{
Large Scale Galaxy Nearest Neighbors$^a$}
\begin{flushleft}
\begin{tabular}{llrrllrr}
\hline
{\bf Galaxy} & Characteristic & V$_{sys}^b$ & V$_{\sigma}^c$ & Companion$^d$ & Hubble & V$^e_{sys}$ & Separation $^f$ \cr
{\bf NGC/IC} & & km s$^{-1}$ & km s$^{-1}$ &  & type & km s$^{-1}$ & kpc \cr
\hline
N 0672  & ND & 408 & 57 & I1727 & SBm(s) & 338 & 18\cr
        &    &     &    & [HKK97] L010 & Irr & 368 & 90\cr
	&    &     &    & LE166062 & Irr & 420 & 95\cr
	&    &     &    & ED F476-08 & Irr  & 359 & 216\cr
N 1784  & ND & 2316 & - & F0523 & Sd & 2358 & 278 \cr
N 1832  & ND & 1938 & 58 & M-02-14-002 & SBdm(s) & 1970 & 361 \cr
N 2217  & ND & 1612 & 93 & E426-G001 & SABm(s) & 1804 & 111 \cr
        &    &      &    & UA126 & Sbc & 1696 & 351 \cr
N 2223  & ND & 2718 & -  & E489-G050 & S0 & 2892 & 143 \cr
        &    &      &    & N2216 & SABab(r) & 2871 & 610 \cr
N 2336  & ND & 2204 & 111& I0467 & SABc(s)  & 2042 & 199 \cr
        &    &      &    & U03604& S        & 2158 & 753 \cr
	&    &      &    & U04103 & Sd      & 2133 & 1000 \cr
	&    &      &    & U03671 & dIrr    & 2292 & 1140 \cr
N 2339  & ND & 2261 &    & 2MASXJ07$^g$ & & 2256 & 62 \cr
N 2366  & ND & 100  & 108 & N2363 & Irr & 70 & 0.8 \cr
N 2545  & ND & 3384 &    &        &     &    &     \cr
N 2835  & ND & 890  & 68 & UA162  & IBm(s) & 850 & 157 \cr
N 2935  & ND & 2276 & 168 &       &        &     &     \cr
N 3287  & ND & 1307 & 124 & N3301 & RSBa(rs) & 1321 & 197 \cr
N 3504  & ND & 1540 & 95  & N3512 & SABc(rs) & 1376 & 92 \cr
N 3513  & ND & 1194 & 62  & N3511 & SABc(s)  & 1106 & 53 \cr
N 3686  & ND & 1157 & 220 & N3684 & SAbc(rs) & 1163 & 95 \cr
        &    &      &     & N3691 & SBb      & 1085 & 131 \cr
        &    &      &     & N3681 & SABbc(r) & 1239 & 191 \cr
	&    &      &     & LSBC D570-02 & Im & 1208 & 270 \cr
	&    &      &     & LSBC D570-01 & Sm & 1019 & 286 \cr
	&    &      &     & [RC3]112$^g$ & Im & 1067 & 388 \cr
\hline
\end{tabular}
\end{flushleft}
\end{table}
\clearpage
\newpage
\setcounter{table}{1}
\begin{table}
\small
\caption[ ]{
Large Scale Galaxy Nearest Neighbors$^a$}
\begin{flushleft}
\begin{tabular}{llrrllrr}
\hline
{\bf Galaxy} & Characteristic & V$_{sys}^b$ & V$_{\sigma}^c$ & Companion$^d$ & Hubble & V$^e_{sys}$ & Separation $^f$ \cr
{\bf NGC/IC} & & km s$^{-1}$ & km s$^{-1}$ &  & type & km s$^{-1}$ & kpc \cr
\hline
N 3729  & ND & 1064 & 148 & N3718 & SBa(s) & 993 & 57\cr
N 3912  & ND & 1789 &   0 & US:U422 &      & 1824 & 153\cr
        &    &      &     & U06791  & Scd  & 1852 & 159\cr
N 3953  & ND & 1047 & 148 & U06840  & SBm(rs) & 1046 & 101\cr
        &    &      &     & N3917A  & SA0     &  837 & 152\cr
	&    &      &     & N3917   & SAcd    &  965 & 204\cr
	&    &      &     & U06802  & Sd      & 1256 & 219\cr
	&    &      &     & SB1147$^g$ &     & 1257 & 250\cr
	&    &      &     & U06983  & SBcd(rs) & 1082 & 265\cr
	&    &      &     & U06923  & Im      & 1062 & 281\cr
	&    &      &     & U06940  & Scd     & 1118 & 322\cr
	&    &      &     & N3992   & SBbc(rs) & 1048 & 353\cr
	&    &      &     & U06969  & Im      &  1119 & 393\cr
	&    &      &     & U06956  & SBm(s)  &   917 & 468\cr
	&    &      &     & U06922  & Scd     &   877 & 469\cr
	&    &      &     & N4026   & S0      &   930 & 479\cr
	&    &      &     & U06667  & Scd     &   973 & 563\cr
	&    &      &     & U06917  & SBm     &   911 & 576\cr
	&    &      &     & N4102   & SABb(s) &   846 & 577\cr
	&    &      &     & N3922   & S0/a    &   906 & 655\cr
	&    &      &     & U06849  & Sm      &   995 & 681\cr
\hline
\end{tabular}
\end{flushleft}
\end{table}
\clearpage
\newpage
\setcounter{table}{1}
\begin{table}
\small
\caption[ ]{
Large Scale Galaxy Nearest Neighbors$^a$}
\begin{flushleft}
\begin{tabular}{llrrllrr}
\hline
{\bf Galaxy} & Characteristic & V$_{sys}^b$ & V$_{\sigma}^c$ & Companion$^d$ & Hubble & V$^e_{sys}$ & Separation $^f$ \cr
{\bf NGC/IC} & & km s$^{-1}$ & km s$^{-1}$ &  & type & km s$^{-1}$ & kpc \cr
\hline
N 3992  & ND & 1053 & 148 & U06940 & Scd & 1118 & 43\cr
        &    &      &     & U06969 & Im  & 1119 & 54\cr
	&    &      &     & U06923 & Im  & 1062 & 72\cr
	&    &      &     & U06983 & SBcd(rs) & 1082 & 209\cr
	&    &      &     & N3953  & SBbc(r)  & 1047 & 354\cr
	&    &      &     & U06894 & Scd      &  849 & 392\cr
	&    &      &     & N4102  & SABb(s)  &  846 & 438\cr
	&    &      &     & U06840 & SBm(rs)  & 1046 & 449\cr
	&    &      &     & N3917A & SA0      &  837 & 498\cr
	&    &      &     & N3982  & SABb(r)  & 1109 & 521\cr
	&    &      &     & N4142  & SBd(s)   & 1157 & 534\cr
	&    &      &     & N3917  & SAcd     &  965 & 553\cr
	&    &      &     & U06802 & Sd       & 1256 & 562\cr
	&    &      &     & N3972  & SAbc(s)  &  852 & 582\cr
	&    &      &     & SB1147$^g$ &      & 1257 & 597\cr
	&    &      &     & N3998  & SA0(r)   & 1040 & 616\cr
	&    &      &     & N3913  & RSAd(rs) &  954 & 659\cr
	&    &      &     & U06919 & Sdm      & 1357 & 670\cr
	&    &      &     & N4026  & S0       &  930 & 720\cr
	&    &      &     & U06956 & SBm(s)   &  917 & 730\cr
	&    &      &     & SB1211$^g$ &      &  907 & 732\cr
N 4123  & ND & 1327 &   0 & N4116  & SBdm(rs) & 1309 & 102\cr
        &    &      &     & U07185 & SAm(rs)  & 1296 & 359\cr
	&    &      &     & U07178 & IABm(rs) & 1339 & 496\cr
	&    &      &     & U07035 & SBa(r)   & 1232 & 507\cr
\hline
\end{tabular}
\end{flushleft}
\end{table}
\clearpage
\newpage
\setcounter{table}{1}
\begin{table}
\small
\caption[ ]{
Large Scale Galaxy Nearest Neighbors$^a$}
\begin{flushleft}
\begin{tabular}{llrrllrr}
\hline
{\bf Galaxy} & Characteristic & V$_{sys}^b$ & V$_{\sigma}^c$ & Companion$^d$ & Hubble & V$^e_{sys}$
& Separation $^f$ \cr
{\bf NGC/IC} & & km s$^{-1}$ & km s$^{-1}$ &  & type & km s$^{-1}$ & kpc \cr
\hline
N 4214  & ND &  291 &  51 & US:U480 NED29  & & 268 & 0.1\cr
        &    &      &     & UA:276  & Im     & 284 & 11\cr
	&    &      &     & N4190   & Im     & 228 & 30\cr
	&    &      &     & KUG1207$^g$ & S  & 339 & 70\cr
	&    &      &     & N4244   & SAcd(s) & 244 & 93\cr
	&    &      &     & U07559  & IBm     & 218 & 149\cr
	&    &      &     & U07605  & Im      & 309 & 165\cr
	&    &      &     & U07599  & Sm      & 278 & 167\cr
N 4236  & ND &    5 &     & IRAS1214$^g$ &    &  95 & 2\cr
        &    &      &     & U07242  & Scd     &  68 & 130\cr
	&    &      &     & U08201  & Im      &  37 & 187\cr
N 4242  & ND &  518 &  58 & U07408  & IAm     & 462 & 89\cr
        &    &      &     & U07320  & dS0     & 521 & 106\cr
	&    &      &     & N4288   & SBdm(s) & 535 & 113\cr
N 4385  & ND & 2150 &  55 & DDO121  & IBm(s)  & 2061 & 219\cr
N 4435  & ND &  725 & 715 & I3355   & Im      &  162 & 64\cr
        &    &      &     & VCC0967 & dE4     & 1135 & 76\cr
	&    &      &     & I3393   & dE      &  476 & 87\cr
\hline
\end{tabular}
\end{flushleft}
\end{table}
\clearpage
\newpage
\setcounter{table}{1}
\begin{table}
\small
\caption[ ]{
Large Scale Galaxy Nearest Neighbors$^a$}
\begin{flushleft}
\begin{tabular}{llrrllrr}
\hline
{\bf Galaxy} & Characteristic & V$_{sys}^b$ & V$_{\sigma}^c$ & Companion$^d$ & Hubble & V$^e_{sys}$
& Separation $^f$ \cr
{\bf NGC/IC} & & km s$^{-1}$ & km s$^{-1}$ &  & type & km s$^{-1}$ & kpc \cr
\hline
	&    &      &     & N4458   & E0      &  635 & 103\cr
	&    &      &     & N4402   & Sb      &  232 & 110\cr
	&    &      &     & VCC0916 & dE      & 1115 & 126\cr
	&    &      &     & VCC0872 & dE0     & 1183 & 128\cr
	&    &      &     & VCC0833 & dE0     &  720 & 138\cr
	&    &      &     & VCC0854 & dE8     &  684 & 153\cr
	&    &      &     & I3363   & dE7     &  790 & 158\cr
	&    &      &     & N4387   & E5      &  561 & 161\cr
	&    &      &     & VCC1001 & I       &  338 & 187\cr
	&    &      &     & VCC0765 & dE1     &  854 & 191\cr
	&    &      &     & I3349   & dE1     & 1563 & 192\cr
	&    &      &     & M84     & E1      & 1060 & 194\cr
	&    &      &     & M87     & E0      &  500 & 223\cr
	&    &      &     & N4436   & dE6     & 1124 & 224\cr
	&    &      &     & N4440   & SBa(rs) &  724 & 230\cr
	&    &      &     & N4431   & SA0(r)  &  934 & 231\cr
	&    &      &     & N4477   & SB0(s)  & 1355 & 234\cr
	&    &      &     & ARK363  & dS0     &  125 & 235\cr
	&    &      &     & N4479   & SB0(s)  &  876 & 237\cr
	&    &      &     & VCC0823 & dE      & 1691 & 267\cr
	&    &      &     & N4486B  & cE0     & 1555 & 267\cr
	&    &      &     & VCC0684 & dE      &  537 & 270\cr
	&    &      &     & N4476   & SA0(r)  & 1970 & 271\cr
\hline
\end{tabular}
\end{flushleft}
\end{table}
\clearpage
\newpage
\setcounter{table}{1}
\begin{table}
\small
\caption[ ]{
Large Scale Galaxy Nearest Neighbors$^a$}
\begin{flushleft}
\begin{tabular}{llrrllrr}
\hline
{\bf Galaxy} & Characteristic & V$_{sys}^b$ & V$_{\sigma}^c$ & Companion$^d$ & Hubble & V$^e_{sys}$
& Separation $^f$ \cr
{\bf NGC/IC} & & km s$^{-1}$ & km s$^{-1}$ &  & type & km s$^{-1}$ & kpc \cr
\hline
N 4487 & ND & 1037 & 106 & N4504 & SAcd(s) & 1003 & 200\cr
       &    &      &     & UA289 & SABdm(s) & 991 & 396\cr
N 4496A & ND & 1732 & 124 & VCC0985 & BCD & 1638 & 260\cr
        &    &      &     & I3474   & Sd  & 1727 & 297\cr
N 4535  & ND & 1956 & 715 & VCC1514 & dE7 & 532  & 109\cr
        &    &      &     & U07688  & Im  & 609  & 138\cr
	&    &      &     & N4519A  & dS0 & 1434 & 159\cr
	&    &      &     & VCC1455 & I   & 1340 & 163\cr
	&    &      &     & VCC1675 & S   & 1795 & 168\cr
	&    &      &     & N4492   & SAa(s) & 1774 & 245\cr
	&    &      &     & N4488   & SB0(s) &  980 & 257\cr
	&    &      &     & VCC1725 & Sm III & 1067 & 265\cr
	&    &      &     & N4522   & SBcd(s) & 2330 & 291\cr
	&    &      &     & VCC1596 & I      & 1286 & 294\cr
	&    &      &     & U07802  & Sdm    & 1788 & 305\cr
	&    &      &     & I3521   & SBm pec & 595 & 306\cr
	&    &      &     & VIII ZW 189 &    & 1350 & 311\cr
	&    &      &     & UA284   & E      & 1155 & 314\cr
	&    &      &     & N4570   & S0(7)/E & 1730 & 335\cr
	&    &      &     & M49     & E2     &  997 & 336\cr
	&    &      &     & VCC1199 & E2     &  900 & 347\cr
	&    &      &     & N4471   & E      &  809 & 351\cr
	&    &      &     & N4467   & E2     &  1426 & 356\cr
	&    &      &     & N4483   & SB0(s) &   875 & 358\cr
\hline
\end{tabular}
\end{flushleft}
\end{table}
\clearpage
\newpage
\setcounter{table}{1}
\begin{table}
\small
\caption[ ]{
Large Scale Galaxy Nearest Neighbors$^a$}
\begin{flushleft}
\begin{tabular}{llrrllrr}
\hline
{\bf Galaxy} & Characteristic & V$_{sys}^b$ & V$_{\sigma}^c$ & Companion$^d$ & Hubble & V$^e_{sys}$
& Separation $^f$ \cr
{\bf NGC/IC} & & km s$^{-1}$ & km s$^{-1}$ &  & type & km s$^{-1}$ & kpc \cr
\hline
	&    &      &     & N4470   & Sa     &  2340 & 359\cr
	&    &      &     & I3487   & E6     &  1157 & 361\cr
	&    &      &     & N4464   & E3     &  1243 & 362\cr
	&    &      &     & I3617   & SBm III & 2088 & 375\cr
	&    &      &     & N4466   & Sab     &  753 & 380\cr
	&    &      &     & I3430   & Im III  & 2015 & 393\cr
	&    &      &     & I3518   & dS0     & 1440 & 418\cr
	&    &      &     & I3591   & SBm III & 1632 & 422\cr
	&    &      &     & N4598   & SB0     & 1961 & 429\cr
	&    &      &     & VCC1357 & I       &  603 & 431\cr
	&    &      &     & U07596  & Im      &  560 & 438\cr
	&    &      &     & N4578   & SA0(r)  & 2273 & 459\cr
	&    &      &     & U07590  & Sbc     & 1117 & 464\cr
	&    &      &     & U07580  & S0(4)   &  716 & 467\cr
	&    &      &     & N4434   & E0      & 1071 & 488\cr
	&    &      &     & I3576   & Sm      & 1075 & 491\cr
	&    &      &     & N4532   & IBm     & 2012 & 507\cr
	&    &      &     & N4451   & Sbc     &  864 & 515\cr
	&    &      &     & VCC1804 & Im      & 1898 & 523\cr
	&    &      &     & I3562   & Im III  & 2051 & 523\cr
	&    &      &     & U07567  & Im      &  867 & 524\cr
	&    &      &     & VCC1164 & E       & 1040 & 525\cr
	&    &      &     & VCC1141 & BCD     & 1040 & 532\cr
	&    &      &     & VCC0989 & S       & 1846 & 535\cr
\hline
\end{tabular}
\end{flushleft}
\end{table}
\clearpage
\newpage
\setcounter{table}{1}
\begin{table}
\small
\caption[ ]{
Large Scale Galaxy Nearest Neighbors$^a$}
\begin{flushleft}
\begin{tabular}{llrrllrr}
\hline
{\bf Galaxy} & Characteristic & V$_{sys}^b$ & V$_{\sigma}^c$ & Companion$^d$ & Hubble & V$^e_{sys}$
& Separation $^f$ \cr
{\bf NGC/IC} & & km s$^{-1}$ & km s$^{-1}$ &  & type & km s$^{-1}$ & kpc \cr
\hline
	&    &      &     & I3414   & SABdm   &  543 & 548\cr
	&    &      &     & VCC1617 & dS0     & 1607 & 553\cr
	&    &      &     & N4416   & SBcd(rs) & 1392 & 555\cr
	&    &      &     & Holm VII & Im     & 2039 & 557\cr
	&    &      &     & N4415    & S0     &  910 & 560\cr
	&    &      &     & VCC1822  & Im     & 1012 & 581\cr
	&    &      &     & N4411b   & SABcd(s) & 1270 & 583\cr
	&    &      &     & N4612    & RSAB0  & 1875 & 584\cr
	&    &      &     & VCC0888  & I      & 1096 & 584\cr
	&    &      &     & U07557   & Sm     &  933 & 588\cr
	&    &      &     & N4623    & SB0    & 1892 & 590\cr
	&    &      &     & VCC1013  & I      & 1712 & 598\cr
	&    &      &     & N4411    & SBc(rs) & 1281 & 602\cr
	&    &      &     & VCC0867  & I      & 1304 & 603\cr
	&    &      &     & N4543    & E3     & 2464 & 615\cr
	&    &      &     & I3602    & dE6    & 1279 & 617\cr
	&    &      &     & I3468    & E1     & 1288 & 621\cr
	&    &      &     & I3412    & I      &  764 & 636\cr
	&    &      &     & VCC1744  & BCD    & 1150 & 638\cr
	&    &      &     & I3322    & SABcd(s) & 1194 & 641\cr
	&    &      &     & U07854   & dE6      & 1032 & 650\cr
	&    &      &     & N4442    & SB0(s)   &  532 & 654\cr
	&    &      &     & VCC1952  & Im IV    & 1308 & 658\cr
\hline
\end{tabular}
\end{flushleft}
\end{table}
\clearpage
\newpage
\setcounter{table}{1}
\begin{table}
\small
\caption[ ]{
Large Scale Galaxy Nearest Neighbors$^a$}
\begin{flushleft}
\begin{tabular}{llrrllrr}
\hline
{\bf Galaxy} & Characteristic & V$_{sys}^b$ & V$_{\sigma}^c$ & Companion$^d$ & Hubble & V$^e_{sys}$
& Separation $^f$ \cr
{\bf NGC/IC} & & km s$^{-1}$ & km s$^{-1}$ &  & type & km s$^{-1}$ & kpc \cr
\hline
	&    &      &     & VCC1605  & S    & 1077 & 658\cr
	&    &      &     & VCC1661  & dE0  & 1400 & 658\cr
	&    &      &     & VCC1933  & S    & 2409 & 661\cr
	&    &      &     & N4417    & SB0  &  843 & 678\cr
	&    &      &     & VCC1896  & dSB0 & 1731 & 683\cr
	&    &      &     & I3322A   & SBcd(s) & 995 & 690\cr
N 4654  & ND & 1037 & 715 & VCC1993  & E0   & 875  & 57\cr
        &    &      &     & N4639    & SABbc(rs) & 1010 & 85\cr
	&    &      &     & VCC1941  & dE1  & 1213 & 93\cr
	&    &      &     & VCC1931  & I    & 1100 & 99\cr
	&    &      &     & N4659    & S0/a &  510 & 116\cr
	&    &      &     & I3742    & SBc(s) & 963 & 128\cr
	&    &      &     & N4620    & S0   & 1156 & 149\cr
	&    &      &     & I3735    & E    & 1895 & 193\cr
	&    &      &     & I3718    & S    &  849 & 234\cr
	&    &      &     & N4640    & dS0  & 1931 & 255\cr
	&    &      &     & I3635    & dE   & 1517 & 275\cr
	&    &      &     & U07906   & Im IV & 1010 & 296\cr
	&    &      &     & VCC1886  & dE5   & 1159 & 305\cr
	&    &      &     & N4641    & S0    & 2017 & 320\cr
	&    &      &     & N4689    & SAbc(rs) & 1616 & 330\cr
	&    &      &     & 2MASXJ124$^g$ & S0  & 1159 & 305\cr
	&    &      &     & VCC1816  & Im IV    & 1006 & 342\cr
	&    &      &     & N4634    & SBcd     &  297 & 354\cr
\hline
\end{tabular}
\end{flushleft}
\end{table}
\clearpage
\newpage
\setcounter{table}{1}
\begin{table}
\small
\caption[ ]{
Large Scale Galaxy Nearest Neighbors$^a$}
\begin{flushleft}
\begin{tabular}{llrrllrr}
\hline
{\bf Galaxy} & Characteristic & V$_{sys}^b$ & V$_{\sigma}^c$ & Companion$^d$ & Hubble & V$^e_{sys}$
& Separation $^f$ \cr
{\bf NGC/IC} & & km s$^{-1}$ & km s$^{-1}$ &  & type & km s$^{-1}$ & kpc \cr
\hline
	&    &      &     & I3779    & dE5     &  1193 & 372\cr
	&    &      &     & N4633    & SABdm(s) &  291 & 373\cr
	&    &      &     & N4584    & SABa(s) &  1779 & 403\cr
	&    &      &     & N4607    & SBb     &  2257 & 413\cr
	&    &      &     & N4606    & SBa(s)  &  1664 & 415\cr
	&    &      &     & I0809    & E       &   206 & 422\cr
	&    &      &     & N4647    & SABc(rs) & 1422 & 453\cr
	&    &      &     & M59      & E5       &  410 & 454\cr
	&    &      &     & M60      & E2       & 1117 & 461\cr
N 4688  & ND & 981  &     & CGCG043-029 & I     & 1051 & 32\cr
        &    &      &     & CGCG043-030 & I     & 1040 & 108\cr
	&    &      &     & U07983      & Im    &  694 & 210\cr
	&    &      &     & N4701       & SAcd(s) & 723 & 302\cr
N 4902  & ND & 2621 &  74 & N4887       & S0 pec  & 2687 & 117\cr
        &    &      &     & N4899       & SBc(rs) & 2658 & 389\cr
	&    &      &     & PGC045114   & I       & 2612 & 443\cr
	&    &      &     & PGC045101   & I       & 2555 & 440\cr
N 4981  & ND & 1677 & 106 & I4212       & SBcd(s) & 1485 & 402\cr
N 5068  & ND &  671 &     &             &         &      &    \cr
N 5792  & ND & 1924 &     & CGCG020-040 &         & 1901 & 218\cr
        &    &      &     & U09601      & SBcd(s) & 1839 & 325\cr
	&    &      &     & U09682      & SBm(s)  & 1815 & 826\cr
	&    &      &     & Mrk1390     & S       & 1756 & 1020\cr
	&    &      &     & 2MASXJ1457$^g$  &         & 1886 & 1080\cr
\hline
\end{tabular}
\end{flushleft}
\end{table}
\clearpage
\newpage
\setcounter{table}{1}
\begin{table}
\small
\caption[ ]{
Large Scale Galaxy Nearest Neighbors$^a$}
\begin{flushleft}
\begin{tabular}{llrrllrr}
\hline
{\bf Galaxy} & Characteristic & V$_{sys}^b$ & V$_{\sigma}^c$ & Companion$^d$ & Hubble & V$^e_{sys}$
& Separation $^f$ \cr
{\bf NGC/IC} & & km s$^{-1}$ & km s$^{-1}$ &  & type & km s$^{-1}$ & kpc \cr
\hline
N 5850  & ND & 2558 & 344 & N5846A      & cE2-3   & 2201 & 83\cr
        &    &      &     & CGCG021-015 &         & 2148 & 130\cr
	&    &      &     & CGCG021-011 & S       & 2035 & 150\cr
	&    &      &     & N5846:62$^g$ &        & 2222 & 168\cr
	&    &      &     & N5846:39$^g$ &        & 2133 & 222\cr
	&    &      &     & N5846:33$^g$ &        & 2303 & 241\cr
	&    &      &     & CGCG021-009  &        & 1970 & 242\cr
	&    &      &     & N5846:28$^g$ &        & 1966 & 326\cr
	&    &      &     & N5846:49$^g$ &        & 2377 & 329\cr
	&    &      &     & N5846:17$^g$ &        & 2002 & 497\cr
	&    &      &     & U09760       & Sd     & 2023 & 614\cr
	&    &      &     & N5869        & S0     & 2087 & 631\cr
	&    &      &     & SD:J1506$^g$ &        & 2019 & 665\cr
N 7640  & ND &  374 &  29 & U12588       & Sdm    &  415 & 103\cr
        &    &      &     & U12632       & Sm     &  422 & 220\cr
N 7723  & ND & 1861 &   0 & MCG0260$^g$  & SBd(rs) & 1944 & 255\cr
        &    &      &     & N7727        & SABa(s) & 1855 & 293\cr
	&    &      &     & N7724        & RSBb(r) & 1927 & 305\cr
\hline
\end{tabular}
\end{flushleft}
\end{table}
\clearpage
\newpage
\setcounter{table}{1}
\begin{table}
\small
\caption[ ]{
Large Scale Galaxy Nearest Neighbors$^a$}
\begin{flushleft}
\begin{tabular}{llrrllrr}
\hline
{\bf Galaxy} & Characteristic & V$_{sys}^b$ & V$_{\sigma}^c$ & Companion$^d$ & Hubble & V$^e_{sys}$
& Separation $^f$ \cr
{\bf NGC/IC} & & km s$^{-1}$ & km s$^{-1}$ &  & type & km s$^{-1}$ & kpc \cr
\hline
N 7741  & ND &  753 &  20 & U12791       & Im      &  792 & 241\cr
\hline
\end{tabular}
\end{flushleft}

$^a$ Extracted from NED (angular separation and systemic velocity) and T88 (members of groups).

$^b$ Systemic velocity from HI observations (\cite{huc89}).

$^c$ Probable velocity dispersion of galaxies in cloud or group (\cite{tul87}).

$^d$ Nomenclature according to catalog,  E:ESO; ED:ESOD; F:FGC; I:IC; LE: LEDA; M:MCG; N:NGC; SB:SBS;
SD:SDSS; U:UGC; UA:UGCA; US: USGC

$^e$ Systemic velocity of companion.

$^f$ Calculated from angular separation (from NED) and distance.

$^g$ IRAS1214: IRAS 12140+6947; KUG1207: KUG 1207+367;
MCG0260: MCG-02-60-010; N5846:XX: NGC 5846[ZM98]00XX; [RC3]112: [RC3]1127.3+1642;
SB1147: SBS 1147+520; SB1211: SBS 1211+540; SD:J143: SDSS J143950.03-004222.9;
SD:J1506: SDSS J150634.25+001255.7; 2MASXJ07: 2MASX J07084902+1845271;
2MASXJ124: 2MASX J12412460+1210328; 2MASXJ1457: 2MASX J14575308+0056035;

$^h$ Separation estimated from angular separation on the plane of the sky (from NED) but without
redshift or systemic velocity information of NGC 6908; alignment is probably a chance superposition
since the H$\alpha$ image from NGC 6907 shows no emission from NGC 6908 (\cite{gar96}).

\end{table}
\clearpage
\newpage
\begin{table}
\small
\caption[ ]{
Companions within 10 and 20 diameters}
\begin{flushleft}
\begin{tabular}{llcc}
\hline
{\bf Galaxy} & Characteristic & Companions &  Companions \cr
{\bf NGC/IC} & & at d$\leq10$D$^a$ & at 10D$\leq d\leq$20D $^b$ \cr
\hline
N 3185 & Sy 2 & 3 & 1 \cr
N 3367 & Sy 2-like, CNS, PM & 0 & 1 \cr
N 3783 & Sy 1& 0 & 0 \cr
N 4477 & Sy 2 & 4 & 0 \cr
N 4507 & Sy 2 & 0 & 1 \cr
N 4725 & Sy 2 & 3 & 4 \cr
N 5135 & Sy 2, CNS & 2 & 5 \cr
N 5347 & Sy 2, CNS, NA& 0 & 0 \cr
N 5728 & Sy 2, CNS & 0 & 0 \cr
N 6217 & Sy 2, NA  & 0 & 0 \cr
N 6951 & Sy 2, CNS, NA & 0 & 1 \cr
N 7479 & Sy 2 & 0 & 2 \cr
\hline
N 1022 & CNS & 0 & 2 \cr
N 1326 & CNS & 3 & 4 \cr
N 1415 & CNS & 2 & 5 \cr
N 3318 & CNS & 1 & 1 \cr
N 3351 & CNS & 2 & 6 \cr
N 4314 & CNS & 5 & 4 \cr
N 5430 & CNS & 0 & 1 \cr
\hline
N 1637  & PM & 0 & 0 \cr
N 2798  & PM & 2 & 1 \cr
N 3319  & PM & 0 & 0 \cr
N 4618  & PM & 2 & 1 \cr
\hline
\end{tabular}
\end{flushleft}
\end{table}
\clearpage
\newpage
\setcounter{table}{2}
\begin{table}
\small
\caption[ ]{
Companions within 10 and 20 diameters}
\begin{flushleft}
\begin{tabular}{llcc}
\hline
{\bf Galaxy} & Characteristic & Companions  &  Companions \cr
{\bf NGC/IC} & & at d$\leq10$D$^a$ & at 10D$\leq d\leq$20D $^b$ \cr
\hline
N 5534  & PM & 0 & 0 \cr
N 5597  & PM & 1 & 0 \cr
N 5691  & PM & 0 & 3 \cr
N 5757  & PM & 1 & 1 \cr
N 5915  & PM & 2 & 0 \cr
N 6907  & PM & 1 & 1 \cr
\hline
I 1953  & NA, PM & 1 & 2 \cr
N 2525  & NA, PM & 0 & 0 \cr
N 2787  & NA & 0 & 1 \cr
N 3359  & NA & 0 & 0 \cr
N 4561  & NA & 0 & 0 \cr
N 4691  & NA & 0 & 0 \cr
N 5669  & NA & 1 & 0 \cr
N 5921  & NA & 1 & 0 \cr
N 6239  & NA & 0 & 0 \cr
IC 5273 & NA, PM  & 0 & 3 \cr
\hline
N 0672  & ND & 4 & 0 \cr
N 1784  & ND & 1 & 0 \cr
N 1832  & ND & 0 & 1 \cr
N 2217  & ND & 1 & 1 \cr
N 2223  & ND & 1 & 1 \cr
N 2336  & ND & 1 & 3 \cr
N 2339  & ND & 1 & 0 \cr
\hline
\end{tabular}
\end{flushleft}
\end{table}
\clearpage
\newpage
\setcounter{table}{2}
\begin{table}
\small
\caption[ ]{
Companions within 10 and 20 diameters}
\begin{flushleft}
\begin{tabular}{llcc}
\hline
{\bf Galaxy} & Characteristic & Companions  &  Companions \cr
{\bf NGC/IC} & & at d$\leq10$D$^a$ & at 10D$\leq d\leq$20D $^b$ \cr
\hline
N 2366 & ND & 1 & 0\cr
N 2545 & ND & 0 & 0\cr
N 2835 & ND & 1 & 0\cr
N 2935 & ND & 0 & 0\cr
N 3287 & ND & 0 & 1\cr
N 3504 & ND & 1 & 0\cr
N 3513 & ND & 1 & 0\cr
N 3686 & ND & 3 & 3\cr
N 3729 & ND & 1 & 0\cr
N 3912 & ND & 0 & 2\cr
N 3953 & ND & 8 & 10\cr
N 3992 & ND & 5 & 16\cr
N 4123 & ND & 1 & 3\cr
N 4214 & ND & 4 & 4\cr
N 4236 & ND & 2 & 1\cr
N 4242 & ND & 2 & 1\cr
N 4385 & ND & 0 & 1\cr
N 4435 & ND & 7 & 19\cr
N 4487 & ND & 1 & 1\cr
N 4496A & ND & 0 & 2\cr
N 4535 & ND & 16 & 57\cr
N 4654 & ND & 9 & 18\cr
N 4688 & ND & 2 & 2\cr
N 4902 & ND & 1 & 3\cr
\hline
\end{tabular}
\end{flushleft}
\end{table}
\clearpage
\newpage
\setcounter{table}{2}
\begin{table}
\small
\caption[ ]{
Companions within 10 and 20 diameters}
\begin{flushleft}
\begin{tabular}{llcc}
\hline
{\bf Galaxy} & Characteristic & Companions  &  Companions \cr
{\bf NGC/IC} & & at d$\leq10$D$^a$ & at 10D$\leq d\leq$20D$^b$ \cr
\hline
N 4981 & ND & 0 & 1\cr
N 5068 & ND & 0 & 0\cr
N 5792 & ND & 2 & 3\cr
N 5850 & ND & 9 & 4\cr
N 7640 & ND & 2 & 0\cr
N 7723 & ND & 0 & 3\cr
N 7741 & ND & 0 & 1\cr
\hline
\end{tabular}
\end{flushleft}
$^a$ Companions within a distance separation less than or equal to 10 diameters of a given bar galaxy;
diameter of each galaxy and angular distances of companions to that galaxy taken from NED.

$^b$ Companions within a distance separation between 10 and 20 diameters of the given bar
galaxy
\end{table}

\clearpage
\newpage
\begin{table}
\small
\caption[ ]{
Spiral Active Galaxies: comparison}
\begin{flushleft}
\begin{tabular}{llllr}
\hline
{\bf Galaxy}  & Hubble & Characteristic & Tully's   & Distance  \cr
{\bf NGC/IC} & Type    & $^a$& Group$^b$ & Mpc       \cr
\hline
N 1068 & RSAb(rs) & Sy 1, CNS & 52-2+1 & 14.4 \cr
N 1097 & RSBb(r)  & Sy 1, CNS, PM & 51-3+1 & 14.5 \cr
M 82   & S & Starburst & 14-10 & 3.6$^c$ \cr
N 5548 & RSA0(s)  & Sy 1.5 &  & 68.9\cr
\hline
\end{tabular}
\end{flushleft}
$^a$ Main Characteristic: Sy: Seyfert type; CNS: circumnuclear structure; PM: peculiar disk
morphology.

$^b$ T88; first digit specifies one of seven regions in T88, the second digit identifies
a specific cloud within the general region; for a complete description refer to section 1 of this
paper and Tully(1987) 14: Coma I Sculptur cloud;  51: Fornax cluster and Eridanus cloud;
52: Cetus-Aries cloud.

$^c$ Cosmological distance taken similar to M81, 3.63 Mpc (\cite{fre94}).
\end{table}
\clearpage
\newpage
\begin{table}
\small
\caption[ ]{
Nearest Neighbors$^a$ of Active Spiral Galaxies}
\begin{flushleft}
\begin{tabular}{llrrllrr}
\hline
{\bf Galaxy} & Characteristic & V$_{sys}^b$ & V$_{\sigma}^c$ & Companion$^d$  & Hubble & V$^e_{sys}$
& Separation $^f$ \cr
{\bf NGC/IC} & & km s$^{-1}$ & km s$^{-1}$ &  & type & km s$^{-1}$ & kpc \cr
\hline
N 1068 & Sy 1 & 1140 & 75 & 2MASXJ024$^g$ & Irr & 1083 & 43 \cr
         & & & & U02162 & IBm(s) & 1185 & 344 \cr
	 & & & & N1073 & SBc(rs) & 1211 & 356 \cr
N 1097 & Sy 1, CNS, PM & 1278 & 103 & N1097A & E pec & 1368 & 14 \cr
       &               &      &     & E416-032 & Sp  & 1221 & 195\cr
       &               &      &     & LS:F416-012 & S & 1076 & 303\cr
       &               &      &     & N1079       & R1R2SABa(rl) & 1447 & 352\cr
M 82 & Starburst & 184 & 108 & M81 & Sab(s) & -34 & 39 \cr
     &           &     &     & Holm IX & Im & 46 & 41\cr
     &           &     &     & BK 03N  & Im & -40 & 46\cr
     &           &     &     & U05423  & Im & 350 & 67\cr
     &           &     &     & N3077   & I0 pec & 14 & 73\cr
     &           &     &     & The Garland & Im & 50 & 76\cr
     &           &     &     & N2976   & Sc pec &  3 & 121\cr
     &           &     &     & Holm I  & Im(s)  & 143 & 124\cr
     &           &     &     & U05442  & Im     & -18 & 133\cr
     &           &     &     & HIJASSJ10$^g$ &  &  46 & 153\cr
     &           &     &     & U05692  & Sm     & 180 & 194\cr
     &           &     &     & I2574   & SABm(s) & 57 & 199\cr
     &           &     &     & U05428  & Im      & 126 & 204\cr
\hline
\end{tabular}
\end{flushleft}
\end{table}
\clearpage
\newpage
\setcounter{table}{4}
\begin{table}
\small
\caption[ ]{
Nearest Neighbors$^a$ of Active Spiral Galaxies}
\begin{flushleft}
\begin{tabular}{llrrllrr}
\hline
{\bf Galaxy} & Characteristic & V$_{sys}^b$ & V$_{\sigma}^c$ & Companion$^d$  & Hubble & V$^e_{sys}$
& Separation $^f$ \cr
{\bf NGC/IC} & & km s$^{-1}$ & km s$^{-1}$ &  & type & km s$^{-1}$ & kpc \cr
\hline
N 5548 & Sy 1.5, PM & 5165 &   & U09165 & S & 5259 & 322\cr
         & & & & CGCG133-031 & Sbc & 5358 & 365 \cr
	 & & & & N5559 & SBb & 5166 & 525 \cr
\hline
\end{tabular}
\end{flushleft}
$^a$ Extracted from NED (angular separation and systemic velocity) and T88 (members of groups).

$^b$ Systemic velocity from HI observations (\cite{huc89}).

$^c$ Probable velocity dispersion of galaxies in cloud or group (\cite{tul87}).

$^d$ Nomenclature according to catalog,  E:ESO; I:IC; N:NGC; U:UGC.

$^e$ Systemic velocity of companion.

$^f$ Calculated from angular separation (from NED) and distance.

$^g$ HIJASSJ10: HIJASS J1021+6842; MASXJ024: 2MASX J02420036+0000531;
\end{table}
\begin{table}
\small
\caption[ ]{
Companions within 10 and 20 diameters of spiral galaxies with known central activity}
\begin{flushleft}
\begin{tabular}{llcc}
\hline
{\bf Galaxy} & Characteristic & Companions &  Companions \cr
{\bf NGC/IC} & & at d$\leq10$D$^a$ & at 10D$\leq d\leq$20D $^b$ \cr
\hline
N 1068 & Sy 1 & 1 & 2 \cr
N 1097 & Sy 1, CNS, PM & 1 & 3 \cr
M 82 & Starburst & 6 & 7 \cr
N 5548 & Sy 1.5, PM & 0 & 3 \cr
\hline
\end{tabular}
\end{flushleft}

$^a$ Companions within a distance separation less than or equal to 10 diameters of the given bar
galaxy.

$^b$ Companions within a distance separation between 10 and 20 diameters of the given active
galaxy

\end{table}
\clearpage
\newpage

\begin{figure}[t]
\plotfiddle{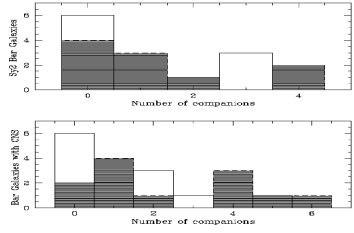}{100pt}{0}{30}{30}{-108}{-36}
\caption[]{Histograms of number of bar galaxies being of Seyfert 2 type (Sy 2) and bar galaxies
with circumnuclear structures (CNS) and their number of companions within 10 diameters (unfilled
rectangles) and within a distance between 10 and 20 diameters (filled rectangles).
}
\label{fig1}
\end{figure}


\begin{figure}[bh]
\plotfiddle{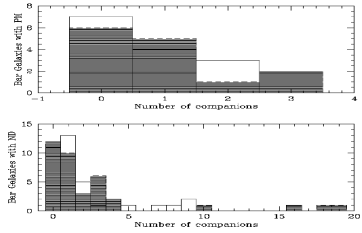}{100pt}{0}{30}{30}{-108}{-72}
\caption[]{Same as in Figure 1 but for bar galaxies with peculiar disk morphology (PM) and
normal disk (ND).
}
\label{fig2}
\end{figure}

\begin{figure}[bh]
\plotfiddle{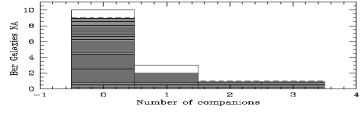}{100pt}{0}{30}{30}{-108}{72}
\caption[]{Same as in Figure 1 but for bar galaxies not associated with
any large scale cloud of galaxies (NA).
}
\label{fig3}
\end{figure}


\clearpage
\newpage

\begin{thebibliography}{99}

\bibitem[Athanassoula 1992]{ath92}Athanassoula, E. 1992, \mnras, 259, 345
\bibitem[Athanassoula 1994]{ath94}Athanassoula, E. 1994 in Mass Trnasfer and Induced Activity in
Galaxies, Ed. I. Shlosman, (Cambridge: Cambridge University Press), p. 143
\bibitem[Bergvall, Laurikainen \& Aalto]{ber03}Bergvall, N., Laurikainen, E. \& Aalto, S.
2003, astro-ph/0304384
\bibitem[Binney \& Tremaine 1987]{bin87}Binney, J., \& Tremaine, S. 1987, Galactic Dynamics
(Princeton: Princeton Univ. Press).
\bibitem[Buta \& Combes 1996]{but96}Buta, R. \& Combes, F. 1996, Fund.Cosmic Phys. 17, 95
\bibitem[Combes 1988]{com88}Combes, F. 1988 in Galactic and Extragalactic Star Formation, Eds. R.E. Pudritz \& M. Fich,
(Dordrecht: Kluwer Academic Plublishers)
\bibitem[Combes 2000]{com00}Combes, F. 2000 in ASP Conf.S., 197, Dynamics of Galaxies:from the Early Universe to the
Present, Eds. F. Combes, G.A. Mamon \& V. Charmandaris (San Francisco: ASP), 15
\bibitem[Combes 2003]{com03}Combes, F. 2003 in ASP Conf.S., xxx, Active Galactic Nuclei: from
Central Engine to Host Galaxy, Eds. S. Collin, F. Combes, \& I. Shlosman  (San Francisco: ASP), 111
\bibitem[Dultzin-Hacyan et al. 1999]{dul99}Dultzin-Hacyan, D., Kronggold, Y., Fuentes-Guridi, I. \& Marziani, P. 1999,
\apj, 513, L111
\bibitem[Freeman et al. 1994]{fre94}Freeman, W. L., Hughes, S.M., Madore, B.F., Mould, J.R., Lee, M.G.,
Stetson, P., Kennicutt, R.C., Turner, A., Ferrarese, L., Ford, H., Graham, J.A., Hill, R.,
Hoessel, J.G., Huchra, J. \& Illingworth G.D. 1994, \apj, 427, 628
\bibitem[Garc\'{\i}a-Barreto et al. 1993]{gar93}Garc\'{\i}a-Barreto, J.A., Carrillo, R., Klein, U. \&
Dahlem, M. 1993, Rev.Mex.Astron.Astrof. 25, 31
\bibitem[Garc\'{\i}a-Barreto et al. 1996]{gar96}Garc\'{\i}a-Barreto, J.A., Franco, J., Carrillo, R., Venegas, S.
\& Escalante-Ram\'{\i}rez, B. 1996, Rev.Mex. Astron. Astrofis., 32, 89
\bibitem[Garc\'{\i}a-Barreto et al. 2002]{gar02}Garc\'{\i}a-Barreto, J.A., Franco, J. \& Rudnick, L.
2002, \aj, 123, 1913
\bibitem[Garc\'{\i}a-Barreto et al. 1998]{gar98}Garc\'{\i}a-Barreto, J.A., Rudnick, L., Franco, J. \&
Martos, M. 1998, \aj, 116, 111
\bibitem[Garc\'{\i}a-Burillo, Combes \& Gerin 1991]{gar91}Garc\'{\i}a-Burillo, S., Combes, F. \& Gerin, M.
1991, IAU 146, Dynamics of Galaxies and their Molecular Cloud Distribution, Ed. F. Combes \& F. Casoli,
p. 351
\bibitem[Gonzalez Delgado et al. 2002]{gon02}Gonzalez Delgado, R.M., Arribas, S., Perez, E., Heckman, T. 2002,
astro-ph/0207275
\bibitem[Helou 1986]{hel86}Helou, G. 1986, \apjl, 311, L33
\bibitem[Hernquist \& Spergel 1992]{her92}Hernquist, L. \& Spergel, D.N. 1992, \apj, 399, L117
\bibitem[Hibbard 2000]{hib00}Hibbard, J.E. 2000 in ASP Conf.S., 197, Dynamics of Galaxies:from the Early Universe to the
Present, Eds. F. Combes, G.A. Mamon \& V. Charmandaris (San Francisco: ASP), 285
\bibitem[Ho, Filippenko \& Sargent 1996]{ho96}Ho, L.C., Filippenko, A.V. \& Sargent, W.L.W. 1996,
in ASP Conf.S., 91, Barred Galaxies, Eds. R. Buta, D.A. Croker, \& B.G. Elmegreen (San Francisco: ASP), 188
\bibitem[Hogg \& Roberts 2001]{hog01}Hogg, D.E. \& Roberts, M.S. 2001, in ASP Conf.S., 240, Gas and
Galaxy Evolution, Eds. J. E. Hibbard, M.P. Rupen, \& J.H. van Gorkom (San Francisco: ASP), 859
\bibitem[Huchtmeier \& Richter 1989]{huc89}Huchtmeier, W.K. \& Richter, O.-G. 1989, A General Catalog of HI Observations
(New York: Springer Verlag)
\bibitem[Ibata, Gilmore \& Irwin 1994]{iba94}Ibata, R.A., Gilmore, G. \& Irwin, M.J. 1994, Nature, 370, 194
\bibitem[Ibata et al. 1997]{iba97}Ibata, R.A., Wyse, R.F.G., Gilmore, G., Irwin, M.J. \& Suntzeff, N.B. 1997, \aj, 113, 634
\bibitem[Knapen, Shlosman \& Peletier 2000]{kna00}Knapen, J.H., Shlosman, I. \& Peletier, R.F. 2000, \apj, 529, 93
\bibitem[Laine \& Heller 1999]{lai99}Laine, S. \& Heller, C.H. 1999, \mnras, 308, 557
\bibitem[Laurikainen \& Salo 1995]{lau95}Laurikainen, E. \& Salo, H. 1995, \aap, 293, 683
\bibitem[Leonard et al. 2003]{leo03}Leonard, D.C., Kanbur, S.M., Ngeow, C.C. \& Tanvir, N.R. 2003
astro-ph/0305259
\bibitem[Martini et al. 2003]{mar03}Martini, P., Regan, M. W., Mulchaey, J.S. \& Pogge, R.W.
2003, \apj, preprint doi:10.1086/374685
\bibitem[Mulchaey, Regan \& Kundu 1997]{mul97}Mulchaey, J.S., Regan, M.W. \& Kundu, A. 1997, \apjs,
110, 299
\bibitem[Noguchi 1988]{nog88}Noguchi, M. 1988, \aap, 203, 259
\bibitem[Phinney 1994]{phi94}Phinney, E.S. 1994 in Mass Trnasfer and Induced Activity in
Galaxies, Ed. I. Shlosman, (Cambridge: Cambridge University Press), p. 1
\bibitem[Rots et al. 1990]{rot90}Rots, A., H., Bosma, A., van der Hulst, J.M., Athanassoula, E. \& Crane, P.C. 1990,
\aj, 100, 387
\bibitem[Sakamoto et al. 1999]{sak99}Sakamoto, K., Okamura, S.K., Ishizuki, S. \& Scoville, N.Z.
1999, \apj, 525, 691.
\bibitem[Sakamoto, Baker \& Scoville 2000]{sak00}Sakamoto, K., Blake, A. J. \& Scoville, N.Z.
2000, \apj, 533, 161.
\bibitem[Salo 1991]{sal91}Salo, H. 1991, \aap, 243, 118
\bibitem[Salo \& Laurikainen 1999a]{sal99a}Salo, H. \& Laurikainen, E. 1999, Ap.Sp.Sc., 269, 589
\bibitem[Salo \& Laurikainen 1999b]{sal99b}Salo, H. \& Laurikainen, E. 1999, Ap.Sp.Sc., 269, 663
\bibitem[Sanders \& Mirabel 1996]{san96}Sanders, D.B. \& Mirabel, I.F. 1996, \araa, 34, 769
\bibitem[Shlosman et al. 1989]{shl89}Shlosman, I., Frank, J. \& Begelman, M.C. 1989, Nature, 338, 45
\bibitem[Toomre 1977]{too77}Toomre, A. 1977 in The Evolution of Galaxies and Stellar Populations, eds. B.M. Tinsley
\& R.B. Larson (Yale Observatory), p. 401
\bibitem[Toomre \& Toomre 1972]{too72}Toomre, A. \& Toomre, J. 1972, \apj, 178, 623
\bibitem[Tully 1987]{tul87}Tully, R.B. 1987, \apj, 321, 280
\bibitem[Tully 1988]{tul88}Tully, R.B. 1988 {\it Nearby Galaxies Catalog}, T88, (Cambridge: Cambridge
Univ. Press).
\bibitem[Tyson et al. 1998]{tys98}Tyson, J.A., Fischer, P., Guhathakurta, P., et al. 1998, \aj, 116,
102
\bibitem[van den Bergh 2002]{van02}van den Bergh, S. 2002, \aj, 124, 782
\bibitem[Walker, Mihos \& Hernquist 1996]{wal96}Walker, I.R., Mihos, J.C. \& Hernquist, L. 1996,
\apj, 460, 121
\bibitem[Waller et al. 1997]{wal97}Waller, W.H., Bohlin, R.C., Cornett, R.H. et al. 1997, \apj, 481,
169
\bibitem[Wada \& Habe 1995]{wad95}Wada, K. \& Habe, A. 1995, \mnras, 277, 433
\bibitem[Yun, Ho \& Lo 1994]{yun94}Yun, M.S., Ho, P.T.P. \& Lo, K.Y. 1994, Nature, 372, 530

\end{thebibliography}
\end{document}